\documentclass{article}
\usepackage{spconf,amsmath,graphicx,hyperref}
\usepackage{amsmath}

\usepackage{xcolor}
\usepackage{hyperref}
\usepackage{amsmath,amssymb,amsfonts}
\usepackage{algorithm}
\usepackage{algorithmicx}
\usepackage{algpseudocode}  
\usepackage{comment}
\usepackage{textcomp}
\usepackage{xcolor}
\usepackage{lscape,lipsum}
\usepackage{threeparttable}
\usepackage{longtable}
\usepackage{multirow}
\usepackage{makecell}
\usepackage{etoolbox}
\usepackage{pgfplots} 
\usepackage{tikz}
\usepackage{graphicx, adjustbox}
\usepackage{subfig}
\usepackage{mathtools}
\usepackage{array}
\usepackage{booktabs}
\usepackage[usestackEOL]{stackengine}
\usepackage{xfrac}
\usepackage{threeparttable}
\usetikzlibrary{shapes.geometric}
\usetikzlibrary{3d}
\usepackage{url}
\usepackage{mathtools}
\usepackage{siunitx}

\usepackage{xcolor}
\usepackage{colortbl}
\usepackage{multirow}
\usepackage{booktabs}

\title{Shapley Features for Robust Signal Prediction in Tactile Internet}
\name{Mohammad Ali Vahedifar, and Qi Zhang \thanks{This research was supported by the TOAST project, funded by the European Union’s Horizon program under the Marie Skłodowska-Curie Actions Doctoral Network (GA No. 101073465), the Danish Council for Independent Research project eTouch (Grant No. 1127- 00339B), and NordForsk Nordic University Cooperation on Edge Intelligence (Grant No. 168043).\\ E-mails: \{av, qz\}@ece.au.dk. Link to the Code: \href{https://github.com/Ali-Vahedifar/Gaussian-Process-Shapley-Feature-Value-for-Signal-Prediction.git}{GitHub Repository}.}}
\address{DIGIT and Department of Electrical and Computer Engineering, Aarhus University, Denmark.}
\begin{document}
\ninept
\maketitle
\begin{abstract}

The Tactile Internet (TI) requires ultra-low latency and reliable haptic signal transmission, yet packet loss and delay remain unresolved challenges. We present a novel prediction framework that integrates Gaussian Processes (GP) with a ResNet-based Neural Network, where GP acts as an oracle to recover signals lost or heavily delayed. To further optimize performance, we introduce Shapley Feature Values (SFV), a principled feature selection mechanism that isolates the most informative inputs for prediction. This GP+SFV framework achieves 95.72\% accuracy, surpassing the state-of-the-art LeFo method by 11.1\%, while simultaneously relaxing TI’s rigid delay constraints. Beyond accuracy, SFV operates as a modular accelerator: when paired with LeFo, it reduces inference time by 27\%, and when paired with GP, by 72\%. These results establish GP+SFV as both a high-accuracy and high-efficiency solution, paving the way for practical and reliable haptic communications in TI systems.

\end{abstract}
\begin{keywords}
Tactile Internet, Signal Prediction, Gaussian Process, Shapley Value, Haptic Communication 
\end{keywords}
\section{Introduction}
\label{sec:intro}

Tactile Internet (TI) is envisioned to enable a human operator to interact with a remote robot by receiving real-time force feedback that allows for the intuitive execution of complex tasks~\cite{zhang20185genabledtactilerobotic}. However, the TI is plagued by two critical challenges. First, the delay between the human's (operator) command and the robot's (teleoperator) response leads to a perceived lack of transparent control and potential instability. Second, missing signals may cause the signals from both sides to become uncorrelated, thereby failing to accurately capture the subtle statistical properties of the motion on the other side~\cite{LeFo}. 

This paper proposes a novel method to accurately compute the anticipated signal from the other side for improving the responsiveness and transparency of haptic teleoperation in the TI. The system's core is two Neural Networks (NNs) explicitly trained by an oracle \textbf{Gaussian Process (GP)} to predict the other side signals. Our approach employs GP to provide ground-truth estimates for guiding NNs, thereby NNs try to mimic GP. This effectively ensures immediate responses to missed or delayed signals~\cite{Rasmussen2006Gaussian}. It is also worth noting that a GP can function as a standalone prediction model. Subsequently, NNs use \textbf{Jensen-Shannon Divergence (JSD)} as a loss function to ensure that predicted haptic signals on one side statistically match the probability distribution of actual haptic signals of the other side over time~\cite{Cover1967}. Moreover, we introduce Shapley Feature Values (SFV), a feature selection mechanism that identifies the most informative features for prediction. Overall, the NNs guided by a GP and enhanced with SFV relax signal loss or delays.
\begin{figure}[t]
    \centering
\includegraphics[width=\linewidth]{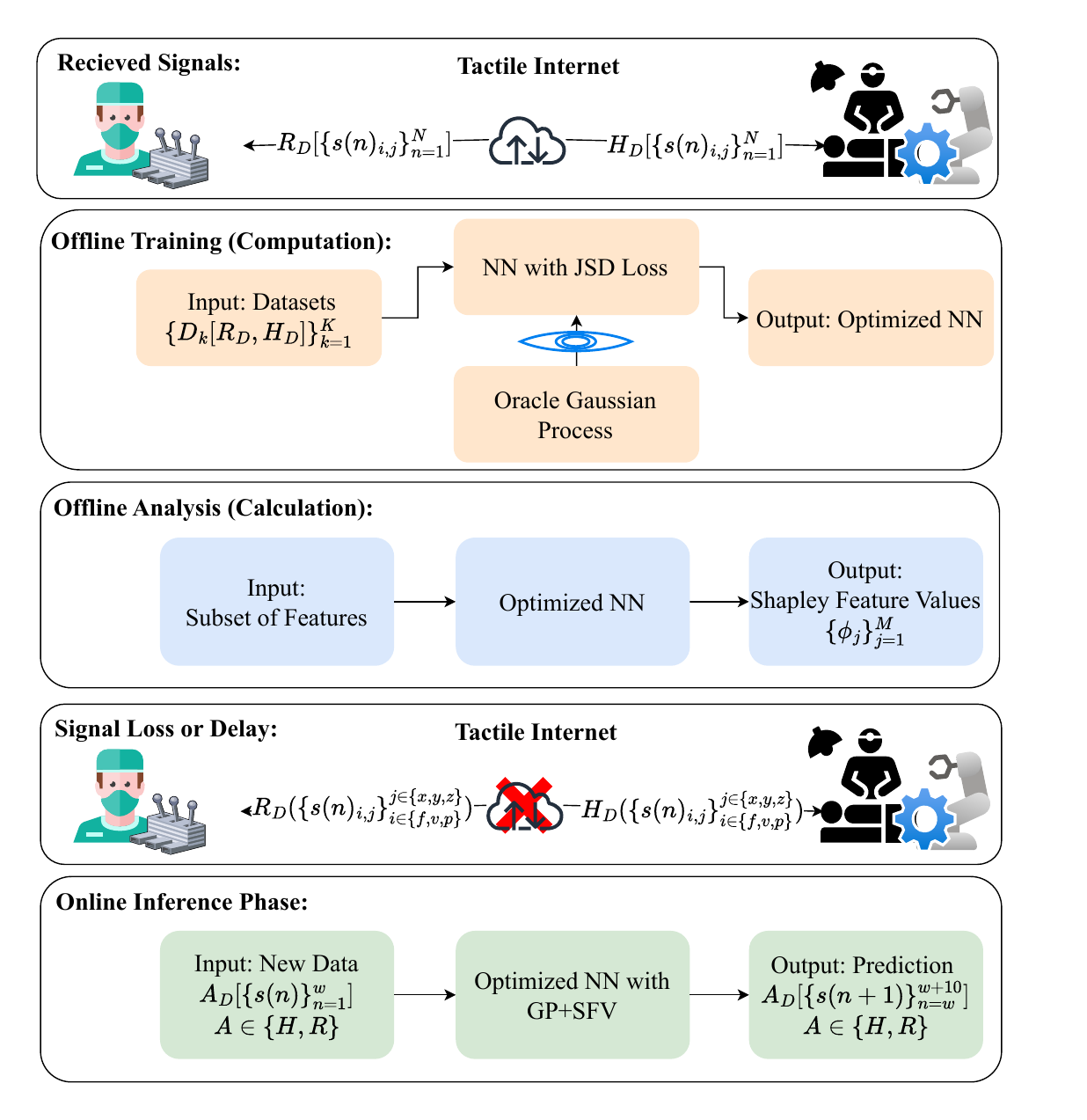}
    \caption{Overview of two-stage predictive framework for TI.}
    \label{fig: Conceptual Fig}
\end{figure}
\section{Related work}
Prior work on haptic signal prediction spans several approaches. Remedy LSTM \cite{xu2020error} improved accuracy over linear estimators but assumed normally distributed data and struggled with missing inputs. A self-attention model \cite{salvato2022predicting} reduced actuator delays but only predicted discrete contact events, not continuous force trajectories. GAN-based augmentation \cite{kizilkaya2023task} introduced synthetic data but raised concerns about fidelity in rare cases. More complex architectures, such as CNN-LSTM-Transformer models \cite{kokkinis2025delayboundrelaxationdeep}, predict force feedback but only in one direction, overlooking bidirectional interaction. Leader-Follower (LeFo)~\cite{LeFo} leverages game-theoretic models for bidirectional haptic signal prediction but has distributional assumptions on haptic data. A key distinction with previous prediction models is that they assume delayed or missing signals must still be received to predict the next sample or use it as a criterion for assessing the prediction. In contrast, GP provides an estimate of the ground truth without requiring such signals. Another distinction is our introduction of a feature selection method to improve prediction performance. Note that the SFV is not limited to our model; it can be applied to any predictive method in TI or signal processing, since it only requires evaluating the importance of features under different setups. This is further illustrated through experiments on LeFo.

\section{System Model}
We consider a teleoperation system in which a human~($H$) operator controls a remote robot~($R$) through bidirectional signal exchange as shown in Fig.~\ref{fig: Conceptual Fig}. Let the human's input signal at time $n$ be a vector $\mathbf{s}(n) = [f(n)_j, v(n)_j, p(n)_j]^T, j\in\{x,y,z\}$ containing force ($f$), velocity ($v$), and position ($p$) values.  We model the relationship between a historical window of these signals, $\{\mathbf{s}(n)_{i,j}\}_{n=1}^{N}, i\in\{f,v,p\}$, and the next signal, $\mathbf{s}(n+1)$, using a two-stage process.

\textbf{i. Training Phase.} First, the historical signals undergo training with the JSD loss function by NNs with GP. The main assumption is that there is no access to ground truth, which represents a key distinction from LeFo, where the challenge involved delayed signals rather than missing ones. Here, GP acts as a probabilistic oracle on the network, learning the underlying distribution of the signals. This stage is offline training (computation).  Second, the relative importance of these features is quantified using SFV. This stage is the offline analysis (calculation). This constitutes a second distinction from LeFo, as our method does not rely on using all available features to predict all features, but instead uses the most important features.

\textbf{ii. Inference Phase:} Once trained, the NNs are optimized for inference. The NNs predict $\mathbf{s}(n+1)$; the criteria for assessing the prediction are GP from offline training. After receiving input samples $n=1$ to $n=10$, it predicts the subsequent samples $n=11$ to $n=20$ by fitting the GP. Prediction accuracy is assessed by comparing NNs outputs against the GP, which serves as a reference distribution. Although the GP does not represent the true distribution exactly, it closely approximates real data. This reliability is further supported by uncertainty quantification, with tolerance levels within 1\% uncertainty guarantee. To maintain robustness, a new GP is fitted after every 10 predicted samples. If actual data for these samples are eventually received with a delay, they are incorporated into the GP; otherwise, the predicted values are used to refit the GP.

\section{Gaussian Process}

We chose GP for this work due to its key advantage in modeling the full posterior distribution of the output. Furthermore, as non-parametric models, GPs offer flexibility to capture complex, non-linear relationships directly from data without assuming a fixed functional form. This makes them well-suited for representing the intricate and diverse movements of human operators and robots.

\noindent The GP is fully specified by its mean function $m(\mathbf{s})$ and its covariance function (kernel) $k(\mathbf{s}, \mathbf{s}')$.
\begin{equation}
f(\mathbf{s}) \sim GP(m(\mathbf{s}), k(\mathbf{s}, \mathbf{s}')).
\end{equation}
Given a history of $n$ training samples, $\mathbf{S} = \{\mathbf{s}_1, \dots, \mathbf{s}_n\}$ with corresponding next states $\mathbf{Y} = \{\mathbf{s}_2, \dots, \mathbf{s}_{n+1}\}$, and a new test input $\mathbf{s}_*$, the predictive distribution for the next state, $\mathbf{s}_{*+1}$, is a Gaussian distribution, which we denote as our target distribution $P_{\mathit{GP}}$:
\begin{equation}
\mathbf{s}_{*+1} | \mathbf{S}, \mathbf{Y}, \mathbf{s}_* \sim P_{\mathit{GP}} = \mathcal{N}(\boldsymbol{\mu}_*, \Sigma_*).
\end{equation}
The predictive mean $\boldsymbol{\mu}_*$ and covariance $\Sigma_*$ are given by:
\begin{align}
&\hspace{-2mm}\boldsymbol{\mu}_* = K(\mathbf{s}_*, \mathbf{S})[K(\mathbf{S}, \mathbf{S}) + \sigma_y^2I]^{-1}\mathbf{Y}, \\
&\hspace{-3mm}\boldsymbol{\Sigma_*} = k(\mathbf{s}_*, \mathbf{s}_*) - K(\mathbf{s}_*, \mathbf{S})[K(\mathbf{S}, \mathbf{S}) + \sigma_y^2I]^{-1}K(\mathbf{S}, \mathbf{s}_*).
\end{align}
where $K(\cdot, \cdot)$ is the kernel matrix and $\sigma_y^2I$ is the noise covariance. This GP's posterior distribution, $P_{GP}$, serves as the reference that our main predictive model will learn to mimic.

\subsection{Neural Network Training with JSD Loss}
Our primary, real-time predictive model is a NN, trained to learn the mapping of the probability distribution parameters, denoted as $Q_{\mathit{NN}} \sim \mathcal{N}(\boldsymbol{\mu}_Q, \boldsymbol{\Sigma}_Q)$. The JSD loss function measures the statistical similarity between the NN's predicted distribution $Q_{\mathit{NN}}$ and the GP's oracle distribution $P_{\mathit{GP}} \sim \mathcal{N}(\boldsymbol{\mu}_P, \boldsymbol{\Sigma}_P)$:
\begin{equation}\label{Eq: JSD}
\mathit{JSD}(P_{\mathit{GP}} \parallel Q_{\mathit{NN}}) = \frac{1}{2} D_{\mathit{KL}}(P_{\mathit{GP}} \parallel M) + \frac{1}{2} D_{\mathit{KL}}(Q_{\mathit{NN}} \parallel M),
\end{equation}
where $M = \frac{1}{2}(P_{\mathit{GP}}+Q_{\mathit{NN}})$ is the mixture distribution. Since $M$ is a two-component Gaussian mixture rather than a single Gaussian, no closed-form solution exists for Eq.~\ref{Eq: JSD}. We therefore employ a Monte Carlo approximation. For reference, the KL divergence between two multivariate Gaussians $\mathcal{N}_1(\boldsymbol{\mu}_1, \boldsymbol{\Sigma}_1)$ and $\mathcal{N}_2(\boldsymbol{\mu}_2, \boldsymbol{\Sigma}_2)$ admits the closed form:
\begin{equation}
\begin{split}
D_{\mathit{KL}}(\mathcal{N}_1 \parallel \mathcal{N}_2) = \frac{1}{2} \Big[ &\mathrm{tr}(\boldsymbol{\Sigma}_2^{-1}\boldsymbol{\Sigma}_1) + (\boldsymbol{\mu}_2 - \boldsymbol{\mu}_1)^T \boldsymbol{\Sigma}_2^{-1} (\boldsymbol{\mu}_2 - \boldsymbol{\mu}_1)\\ 
&- k + \ln \frac{|\boldsymbol{\Sigma}_2|}{|\boldsymbol{\Sigma}_1|} \Big],
\end{split}
\end{equation}
 \noindent where $k$ denotes the dimensionality of the signal. To compute Eq.~\ref{Eq: JSD}, we draw $L$ samples $\{\mathbf{x}^{(l)}\}_{l=1}^{L}$ from the source distribution (e.g., $P_{\mathit{GP}}$) and estimate the divergence as:
\begin{equation}
D_{\mathit{KL}}(P_{\mathit{GP}} \parallel M) \approx \frac{1}{L}\sum_{l=1}^{L} \left[\log p(\mathbf{x}^{(l)}) - \log m(\mathbf{x}^{(i)})\right],
\end{equation}
where $p(\cdot)$ and $m(\cdot)$ denote the densities of $P_{\mathit{GP}}$ and $M$, respectively, with $m(\mathbf{x}) = \frac{1}{2}p(\mathbf{x}) + \frac{1}{2}q(\mathbf{x})$. The term $D_{\mathit{KL}}(Q_{NN} \parallel M)$ is computed analogously. This formulation remains fully differentiable, enabling end-to-end backpropagation.

\section{Shapley Feature Value}
In our framework, we apply the SFV method as an offline analysis of the features' contribution to prediction. Let $\mathcal{P}$ denote the original prediction model used to predict signals in TI, and $\mathcal{E}$ denotes the explanation model that contains all extracted features for prediction. The explanation model tries to ensure:
\begin{equation}
     \mathcal{P}(\mathbf{S})\approx \mathcal{E}(Z), \quad Z=\{z_a\}_{a
=1}^{M}, \quad \mathbf{S}=\{\mathbf{s}_b\}_{b=1}^{n}.
\end{equation}
 Here, $Z$ is the set of extracted features, and $\mathbf{S}$ is the input set for the prediction model. Additive features attribution as an explanation model for a prediction model can be written as:
\begin{equation}
\mathcal{P}(\mathbf{S})\approx\mathcal{E}(Z) = z_0 + \sum_{a=1}^M \phi_a z_a,
\end{equation}
where $z_0 = \mathcal{P}(h_\mathbf{s}(0))$ represents the model output with all features missing (toggled off), $M$ is the number of simplified input features, and $\phi_a \in \mathbb{R}$. Here, $\phi_a$ attributes an importance value to each feature. To find a closed form for the importance value, let us define the sum of total importance values as $V: \phi\rightarrow\phi^M$. We argue that $\phi_a$ should satisfy the following axioms:

\noindent\textbf{1. Transferability:} \label{axiom 1} The total value is distributed among all features:
    \begin{equation}
    \sum_{a \in M} \phi_a = V(Z),
    \end{equation}
    i.e., the sum of individual Shapley values equals the total value.
    
\noindent\textbf{2. Monotonicity:} This axiom satisfies three axioms simultaneously: \textbf{a. Null contribution:} If adding a specific feature does not improve performance, no matter which subset it is added to, then it should have zero value. \textbf{b. Symmetry:} If two features contribute equally to the model, then their effect must be the same. \textbf{c. Linearity:} The effect a feature has on the sum of two
functions is the effect it has on one function, plus the effect it has on the other.
  \begin{equation}
      \begin{split}
          \text{If } V_1(\mathbf{S} \cup \{a\}) - V_1(\mathbf{S}) \geq V_2(\mathbf{S} \cup \{a\}) - V_2(\mathbf{S}),\\
  \text{ for all } \mathbf{S} \subseteq \mathcal{Z} \setminus \{a\}
  \Rightarrow \phi_a(V_1) \geq \phi_a(V_2).
      \end{split}
  \end{equation}
Let us propose the SFV, based on the Shapley value from cooperative GT, which is defined as:
\noindent\newtheorem{theorem}{Theorem}
\begin{theorem}[Shapley Feature Value]
Any Feature valuation $\phi_a$ satisfying Axioms 1 and 2  must have the form:
\begin{equation}\label{Eq:ShapleyValue}
\phi_a = \sum_{\mathbf{S} \subseteq Z \setminus \{a\}} \frac{|\mathbf{S}|!(Z - |\mathbf{S}| - 1)!}{Z!} \left[ \mathcal{E}_{\mathbf{S} \cup \{a\}} - \mathcal{E}_\mathbf{S} \right].
\end{equation}
If and only if we calculate Eq.~\ref{Eq:ShapleyValue} for valuation, then Axioms 1 and 2 are satisfied.
\end{theorem}
Here, \( \mathbf{S} \subseteq Z \setminus \{a\} \) is a subset of features excluding \(a\), $\mathcal{E}_{\mathbf{S} \cup \{a\}}$ is a model trained with including feature $z_a$ in prediction, $\mathcal{E}_\mathbf{S}$ is trained without the feature $z_a$. 
Since the effect of withholding a feature depends on other features in the model, the differences of predictions are computed for all possible subsets $\mathbf{S} \subseteq Z \setminus \{a\}$.  To calculate the Shapley value for a feature $z_a$, one must consider every possible subset of features not containing $z_a$. For each of these subsets, we should calculate the marginal contribution of $z_a$. Generally, computing the exact Shapley value (Eq.~\ref{Eq:ShapleyValue}) involves evaluating $2^M$ subsets, often necessitating approximation methods like Monte Carlo sampling. However, in our system model, the feature space is limited to $M=9$, which renders the exact calculation feasible.

\textbf{Proof.} The Shapley Feature Value derives from the original Shapley Value in cooperative game theory~\cite{shapley:book1952}, which provides the unique fair allocation of rewards among players based on their marginal contributions. Feature valuation can be cast into this framework by treating each feature as a player: combining different feature subsets yields varying prediction performance, analogous to different player coalitions achieving different scores. The SFV thus determines each feature's credit based on its contribution to overall model performance, satisfying the fairness axioms listed above.
\section{Results \& Discussion}
\begin{figure}[t]
    \centering
    \includegraphics[width=\linewidth]{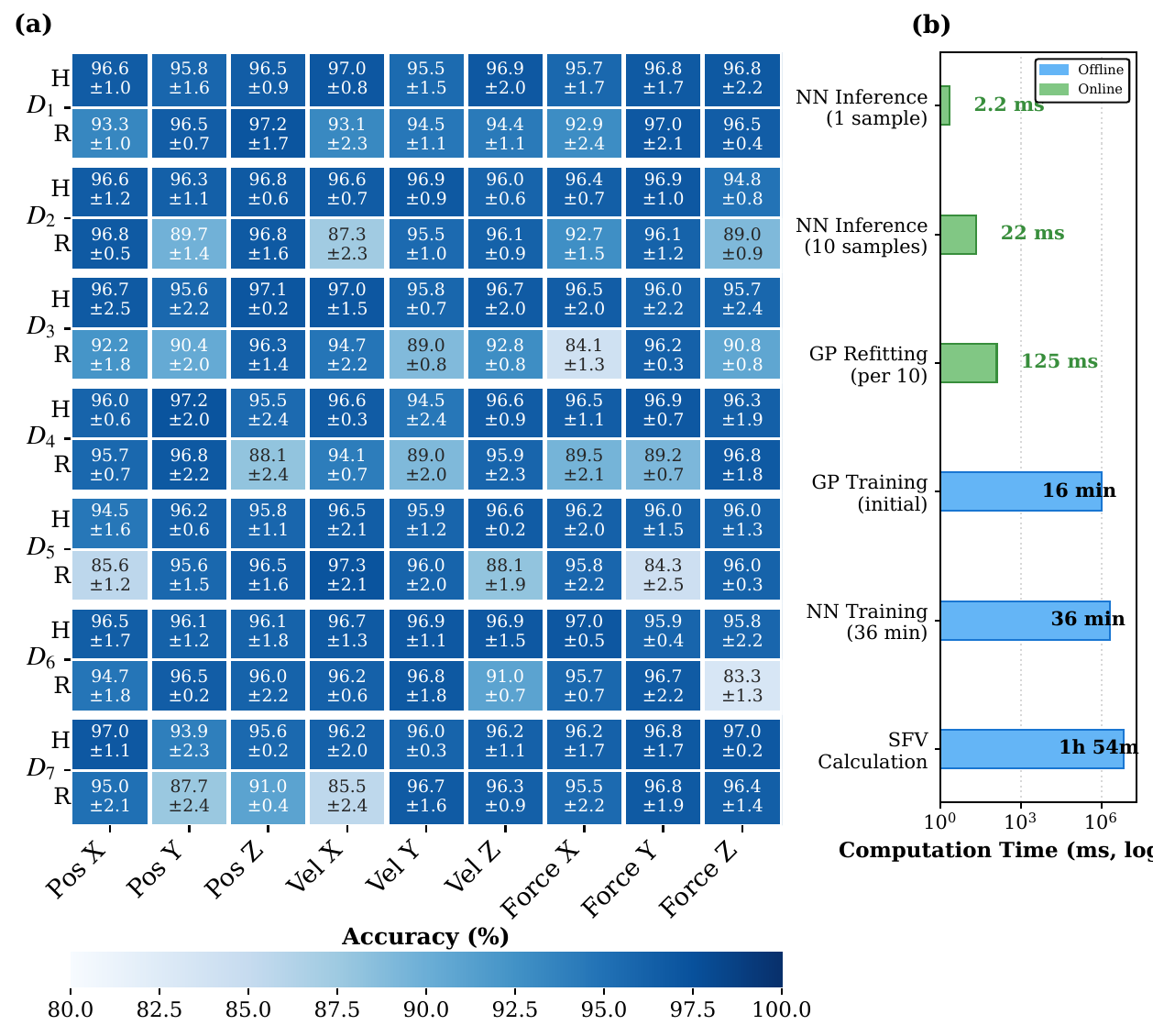}
    \caption{Prediction accuracy and standard deviation for predicting the next 10 samples with ResNet training with GP+SFV.}
    \label{fig: heatmap}
\end{figure}
\begin{table*}[tbp]
\centering
\caption{Accuracy (\%) of $D_1$ Dataset for predicting the next 10 samples. The best results per feature highlighted (green: human, cyan: robot).}
\label{tab:drag_max_stiffness_results}

\renewcommand{\arraystretch}{1.1}
\setlength{\tabcolsep}{4pt}
\scriptsize
\begin{tabular}{l|cc|cc|cc|cc|cc|cc|cc|cc|cc}
\multirow{4}{*}{\textbf{Metric}} & 
\multicolumn{6}{c|}{\textbf{Fully Connected}} & 
\multicolumn{6}{c|}{\textbf{LSTM}} & 
\multicolumn{6}{c}{\textbf{ResNet}} \\
\hline
& \multicolumn{2}{c|}{\textbf{LeFo}} & \multicolumn{2}{c|}{\textbf{LeFo+SFV}} & \multicolumn{2}{c|}{\textbf{GP+SFV}} & \multicolumn{2}{c|}{\textbf{LeFo}} & \multicolumn{2}{c|}{\textbf{LeFo+SFV}} & \multicolumn{2}{c|}{\textbf{GP+SFV}} & \multicolumn{2}{c|}{\textbf{LeFo}} & \multicolumn{2}{c|}{\textbf{LeFo+SFV}} & \multicolumn{2}{c}{\textbf{GP+SFV}} \\
\cline{2-19}
& \textbf{H} & \textbf{R} & \textbf{H} & \textbf{R} & \textbf{H} & \textbf{R} & \textbf{H} & \textbf{R} & \textbf{H} & \textbf{R} & \textbf{H} & \textbf{R} & \textbf{H} & \textbf{R} & \textbf{H} & \textbf{R} & \textbf{H} & \textbf{R} \\
\hline
\textbf{Position X} & 85.22 & 74.71 & 91.36 & 79.76 & 93.88 & 87.51 & 85.88 & 75.18 & 89.14 & 78.96 & 93.01 & 87.78 & 89.15 & 78.60 & 94.63 & 84.57 & \cellcolor{green!15}96.62 & \cellcolor{cyan!15}93.35 \\
\textbf{Position Y} & 93.49 & 88.23 & 95.96 & 94.47 & 95.61 & 95.45 & 94.13 & 89.76 & 95.45 & 93.68 & \cellcolor{green!15}96.34 & 95.10 & 95.02 & 92.56 & 95.92 & \cellcolor{cyan!15}96.73 & 95.82 & 96.49 \\
\textbf{Position Z} & 90.16 & 78.35 & 95.36 & 84.59 & 95.53 & 90.10 & 90.99 & 78.88 & 94.65 & 82.51 & \cellcolor{green!15}95.33 & 89.99 & 93.48 & 81.55 & 95.37 & 87.23 & 96.49 & \cellcolor{cyan!15}97.18 \\
\textbf{Velocity X} & 88.12 & 77.27 & 93.23 & 84.19 & 94.98 & 88.64 & 89.27 & 78.06 & 92.40 & 82.67 & 95.32 & 88.28 & 92.35 & 80.96 & 95.91 & 86.39 & \cellcolor{green!15}97.00 & \cellcolor{cyan!15}93.14 \\
\textbf{Velocity Y} & 81.38 & 74.79 & 88.34 & 80.83 & 91.58 & 90.51 & 81.97 & 75.27 & 86.24 & 79.26 & 92.01 & 84.73 & 85.00 & 79.04 & 91.01 & 84.26 & \cellcolor{green!15}95.46 & \cellcolor{cyan!15}94.47 \\
\textbf{Velocity Z} & 81.40 & 75.00 & 87.67 & 81.53 & 93.63 & 85.11 & 82.81 & 76.54 & 86.83 & 79.24 & 92.80 & 88.21 & 85.66 & 78.24 & 91.64 & 83.56 & \cellcolor{green!15}96.92 & \cellcolor{cyan!15}94.37 \\
\textbf{Force X} & 79.89 & 73.50 & 86.42 & 79.34 & 94.43 & 87.91 & 80.63 & 74.07 & 83.29 & 78.69 & 89.66 & 84.29 & 83.07 & 77.03 & 88.82 & 82.28 & \cellcolor{green!15}95.68 & \cellcolor{cyan!15}92.86 \\
\textbf{Force Y} & 92.20 & 84.98 & 95.71 & 90.83 & 95.63 & 95.52 & 94.19 & 85.45 & 95.49 & 87.98 & 95.46 & 95.32 & 95.90 & 89.38 & \cellcolor{green!15}97.13 & 94.87 & 96.76 & \cellcolor{cyan!15}96.95 \\
\textbf{Force Z} & 88.16 & 79.43 & 95.01 & 85.79 & 95.29 & 95.20 & 89.23 & 80.67 & 92.90 & 83.75 & 95.45 & 91.60 & 91.35 & 83.11 & 95.15 & 89.30 & \cellcolor{green!15}96.81 & \cellcolor{cyan!15}96.55 \\
\hline
\textbf{Average} & 86.67 & 78.47 & 92.12 & 84.59 & 94.51 & 90.66 & 87.68 & 79.32 & 90.71 & 82.97 & 93.93 & 89.48 & 90.11 & 82.27 & 93.95 & 87.69 & \cellcolor{green!15}96.40 & \cellcolor{cyan!15}95.04 \\
\bottomrule
\end{tabular}
\end{table*}

\begin{table}[tbp]
\centering
\caption{Overall result for Drag Max Stiffness Y Dataset.}
\label{tab:performance_summary}
\begin{tabular}{lccc}
\textbf{Architecture} & \textbf{LeFo} & \textbf{LeFo+SFV $\uparrow$} & \textbf{GP+SFV $\uparrow$} \\
\toprule
\textbf{FC} & 82.57 & 88.36 (7.0\%) & 92.59 (12.1\%) \\
\textbf{LSTM} & 83.50 & 86.84 (4.0\%) & 91.71 (9.8\%) \\
\textbf{ResNet} & 86.19 & 90.82 (5.4\%) & \cellcolor{green!15}95.72 (11.1\%) \\
\bottomrule
\end{tabular}
\end{table}

\begin{table}[tbp]
\centering
\caption{Inference Time (ms) for 1 sample prediction.}
\label{tab:inference_comparison_detailed}
\setlength{\tabcolsep}{2pt}
\begin{tabular}{l|ccc|ccc|ccc}
& LeFo & LeFo & \textbf{$\uparrow$} & GP & GP & \textbf{$\uparrow$} & GP vs & GP+SFV vs \\
&&+SFV&&&+SFV&&LeFo$\uparrow$&LeFo+SFV$\uparrow$\\
\hline
$D_1$ & 9.1 & 7.1 & 22\% & 5.9 & \cellcolor{green!15}2.1 & 64\% & 35\% & 70\% \\
$D_2$ & 7.8 & 5.6 & 28\% & 4.5 & \cellcolor{green!15}2.8 & 38\% & 42\% & 50\% \\
$D_3$ & 11.5 & 8.6 & 25\% & 7.1 & \cellcolor{green!15}2.5 & 65\% & 38\% & 71\% \\
$D_4$ & 8.7 & 6.0 & 31\% & 4.7 & \cellcolor{green!15}2.1 & 55\% & 46\% & 65\% \\
$D_5$ & 14.8 & 11.2 & 24\% & 9.9 & \cellcolor{green!15}2.4 & 76\% & 33\% & 79\% \\
$D_6$ & 17.7 & 11.9 & 33\% & 9.2 & \cellcolor{green!15}1.9 & 79\% & 48\% & 84\% \\
$D_7$ & 16.9 & 12.3 & 27\% & 10.0 & \cellcolor{green!15}1.7 & 83\% & 41\% & 86\% \\
\hline
Avg & 12.4 & 8.8 & 27\% & 7.3 & \cellcolor{green!15}2.2 & 66\% & 40\% &  \cellcolor{green!15}72\% \\
\bottomrule
\end{tabular}
\end{table}

\textbf{Datasets:} We utilized real-world haptic data traces in the experiments. The dataset captures kinaesthetic interactions recorded using a Novint Falcon haptic device within a Chai3D virtual environment. The dataset provides detailed records of 3D position, velocity, and force measurements~\cite{rodriguez_guevara_2025_14924062}. Results are averaged over 10 independent runs. Here, $D_1$: Drag Max Stiffness Y, $D_2$: Horizontal Movement Fast, $D_3$: Horizontal Movement Slow, $D_4$: Tap and Hold Max Stiffness Z-Fast, $D_5$: Tap and Hold Max Stiffness Z-Slow, $D_6$: Tapping Max Stiffness Y-Z, $D_7$: Tapping Max Stiffness Z.

\textbf{Neural Network:} We employed three NN architectures: 1.~Fully Connected (FC)~\cite{Goodfellow-et-al-2016}: a 12-layer FC NN, with each layer containing 100 ReLU-activated units. 2.~ResNet-32~\cite{He_2016_CVPR}: network weights initialized using He initialization~\cite{He_2015_ICCV}. 3.~LSTM~\cite{LSTM6795963}:
 two stacked LSTM layers with 128 units each, followed by a dense output layer with a linear activation function.

\textbf{Training:} Training was performed using Dropout~\cite{Dropout10.5555/2627435.2670313} and stochastic gradient descent (SGD) with a momentum of 0.9, an initial learning rate of 0.01, and a batch size of 32.  All datasets were trained with one NVIDIA RTX A6000 GPU. For integrating any component of this work into real-world TI, two steps are essential. First, the model should be pretrained on a dataset specific to the target system, covering diverse movements and conditions. In our case, we used seven datasets with different motions and scenarios to achieve this. Second, during inference, online prediction is performed to produce the results. Deploying our framework in real-world TI applications involves two essential steps: offline pretraining on system-specific datasets encompassing diverse movements and operating conditions (In our case, we used seven datasets), followed by online inference for real-time prediction.

\textbf{Result:} Prediction accuracy heatmaps for different datasets are illustrated in Fig.~\ref{fig: heatmap}.(a), where GP+SFV with a ResNet-based NN achieves nearly 97\% accuracy across multiple features, demonstrating its superior performance\footnote{Heatmaps for other NNs and methods are available in the GitHub link.}. It also highlights that the accuracy of predicting human signals is more robust than that of robot signals across different datasets. In comparison with LeFo, our results demonstrate that treating all features equally is suboptimal. The LeFo paper predicts all available features, without discrimination (i.e., it uses all features to predict all features). In contrast, our approach incorporates SFV to perform a rigorous feature selection process (i.e., most important features to predict all features). In addition, as shown in Fig.~\ref{fig: heatmap}.(b), the computationally expensive operations are performed offline and do not impact real-time performance. Online operations remain fast: GP refitting occurs only every 10 samples (125 ms) and can run asynchronously in the background, while per-sample inference takes just 2.2 ms.
This architecture decouples periodic model updates from real-time prediction, ensuring the system meets teleoperation latency requirements.

As shown in Table~\ref{tab:drag_max_stiffness_results}, GP+SFV consistently outperforms both GP and LeFo across all features and architectures\footnote{Other datasets' results are available at the GitHub link.}. The integration of SFV yields substantial improvements in prediction accuracy, supporting its role in enhancing model performance. GP+SFV also excels in computational efficiency. 
Table~\ref{tab:performance_summary}, GP+SFV achieves the highest overall accuracy, with gains ranging from 11.1\% for ResNet to 12.1\% for the Fully Connected model. These results are reinforced by Table~\ref{tab:drag_max_stiffness_results}, where GP+SFV records the highest prediction accuracy across nearly all signal types and axes, for both the human (H) and robot (R) sides. This demonstrates that the method’s unique training paradigm, where a NN learns from a GP oracle, produces a more robust and statistically aligned predictive model. As in Table~\ref{tab:inference_comparison_detailed}, it achieves the fastest inference across all datasets. For example, in dataset $D_7$, incorporating SFV reduces inference time by 83\% compared to GP alone, and by 86\% compared to LeFo+SFV, highlighting GP’s advantage over LeFo when paired with feature selection. 

\begin{figure}[t]
    \centering
    \includegraphics[width=\linewidth]{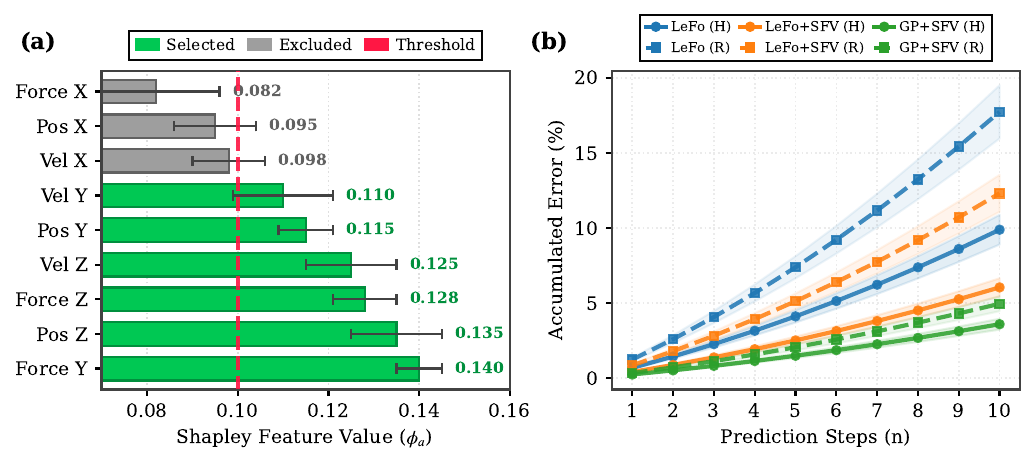}
    \caption{(a) Shapley Feature Value analysis for GP+SFV on ResNet for Drag Max Stiffness Y Dataset. (b) Error prediction of the next 10 samples with ResNet. Solid lines: Human, dashed lines: Robot.}
    \label{fig: error}
\end{figure}

The efficiency gains can be attributed to the principled feature selection enabled by SFV. As illustrated in Fig.~\ref{fig: error}(a), SFV quantifies each feature's contribution to prediction accuracy, revealing that Force Y ($\phi_a=0.140$) and Position Z ($\phi_a=0.135$) are the most critical features for teleoperation performance. By applying a threshold of $\phi_a=0.1$, the framework systematically identifies six high-importance features. This data-driven dimensionality reduction decreases computational cost and inference time without sacrificing accuracy, as the excluded features contribute minimally to predictive power. While the proposed method significantly improves inference efficiency, it is important to examine its behavior over extended prediction horizons. As shown in Fig.~\ref{fig: error}.(b), increasing the number of prediction samples also increases the overall error. This is expected: when the GP becomes outdated for distant samples, it provides inaccurate guidance to the NN, leading to accumulated error. A further observation is that the accumulated error of the robot signals is consistently higher than that of human signals across different methods. This suggests that predicting robots' haptic feedback is more challenging. We attribute this disparity primarily to the inherent characteristics of robot motion. Unlike human movements, which are generally smooth and continuous, robot trajectories often exhibit abrupt transitions caused by discrete control actions and task-specific behaviors such as tapping. These sharp changes are difficult to infer from historical observations, especially when they are not explicitly encoded in the human feature set.
\section{Conclusion}
We propose a two-stage signal prediction framework for TI that integrates the predictive strength of GP as an oracle with the statistical alignment capabilities of the JSD as a loss function. To further enhance performance, we incorporate SFV as a feature selection mechanism. SFV improves both prediction accuracy and computational efficiency by identifying and utilizing the most informative features. Together, these components establish a promising framework for predictive teleoperation systems that balance computational efficiency through SFV with predictive robustness by the GP oracle.

\bibliographystyle{IEEEbib}
\bibliography{strings,refs}
\end{document}